\documentclass[twocolumn,nofootinbib,aps,superscriptaddress,floatfix]{revtex4}

\usepackage{graphicx}
\usepackage{epsfig}
\usepackage{dcolumn}
\newif\ifpdf
\ifx\pdfoutput\undefined
\pdffalse % we are not running PDFLaTeX
\else
\pdfoutput=1 % we are running PDFLaTeX
\pdftrue
\fi

\def\OMIT#1{{}}
\def\lqcd{\ensuremath{\Lambda_{\rm QCD}}}
\def\MSbar{\ensuremath{\overline{\rm MS}}}
\def\GeV{\mbox{GeV}}
\def\MeV{\mbox{MeV}}

\def\mDbar{\overline{m}_D}
\def\mbups{\ensuremath{m_b^{1S}}}

\def\mcms{\ensuremath{\overline{m}_c}}

\def\mbkin{\ensuremath{m_b^{\rm kin}}}
\def\mckin{\ensuremath{m_c^{\rm kin}}}
\def\vev#1{\ensuremath{\left\langle #1 \right\rangle}}

\def\epsBLM{\ensuremath{\epsilon^2_{\rm BLM}}}
\def\d{{\rm d}}

\def\mX#1{\ensuremath{\langle m_X^{#1}\rangle}}

\def\Eg#1{\ensuremath{\langle E_\gamma^{#1}\rangle}}
\def\oneSA{\ensuremath{1S_{\text{EXP}}}}
\def\oneSB{\ensuremath{1S_{\text{NO}}}}
\def\kinA{\ensuremath{\mbox{kin}_{\text{EXP}}}}
\def\kinB{\ensuremath{\mbox{kin}_{\text{NO}}}}
\def\kinC{\ensuremath{\mbox{kin}_{\text{UG}}}}

\def\Ecut{\ensuremath{E_{\rm cut}}}

\def\l#1{\ensuremath{\lambda_{#1}}}
\def\r#1{\ensuremath{\rho_{#1}}}
\def\t#1{\ensuremath{{\cal T}_{#1}}}

\newcommand{\nn}{\nonumber}
\newcommand{\beq}{\begin{equation}}
\newcommand{\eeq}{\end{equation}}
\newcommand{\beqa}{\begin{eqnarray}}
\newcommand{\eeqa}{\end{eqnarray}}

\begin{document}
%%%%%%%%%%%%%%%%%%%%%%%%%%%%%%%%%%%%%%%%%%
%Some more stuff to get graphics to work
\ifpdf
\DeclareGraphicsExtensions{.pdf, .jpg}
\else
\DeclareGraphicsExtensions{.eps, .jpg}
\fi
%%%%%%%%%%%%%%%%%%%%%%%%%%%%%%%%%%%%%%%%%%

\preprint{ \vbox{\hbox{CALT-68-2515} \hbox{LBNL-55945} 
  \hbox{UCSD/PTH 04-12} \hbox{hep-ph/0408002} }}

\title{\boldmath Global analysis of inclusive $B$ decays}
\vspace*{1.5cm}

\author{Christian W.\ Bauer}
\affiliation{California Institute of Technology, Pasadena, CA 91125}

\author{Zoltan Ligeti}
\affiliation{Ernest Orlando Lawrence Berkeley National Laboratory,
  University of California, Berkeley, CA 94720}

\author{Michael~Luke}
\affiliation{Department of Physics, University of Toronto,
  60 St.\ George Street, Toronto, Ontario, Canada M5S 1A7}

\author{Aneesh V.\ Manohar}
\affiliation{Department of Physics, University of California at San Diego,
  La Jolla, CA 92093}

\author{Michael Trott\vspace*{8pt}}
\affiliation{Department of Physics, University of Toronto,
  60 St.\ George Street, Toronto, Ontario, Canada M5S 1A7}

\begin{abstract}
In light of the large amount of new experimental data, we revisit the
determination of $|V_{cb}|$ and $m_b$ from inclusive semileptonic and radiative
$B$ decays.  We study shape variables to order $\lqcd^3/m_b^3$ and
$\alpha_s^2\beta_0$, and include the order $\alpha_s\, \lqcd/m_b$ correction to
the hadron mass spectrum in semileptonic decay, which improves the agreement
with the data.  We focus on the $1S$ and kinetic mass schemes for the $b$ quark,
with and without expanding $m_b-m_c$ in HQET. We perform fits to all available
data from BABAR, BELLE, CDF, CLEO, and DELPHI, discuss the theoretical
uncertainties, and compare with earlier results.  We find $|V_{cb}|  = \left(
41.4 \pm 0.6 \pm 0.1_{\tau_B} \right) \times 10^{-3}$ and $m_b^{1S} = 4.68 \pm
0.03\,\text{GeV}$,  including our estimate of the theoretical uncertainty in the
fit.
\end{abstract}

\maketitle

\section{Introduction}\label{sec:intro}

In the last few years there has been intense theoretical and experimental
activity directed toward a precise determination of the
Cabibbo-Kobayashi-Maskawa (CKM) matrix element $|V_{cb}|$ from combined fits to
inclusive semileptonic $B$ decay
distributions~\cite{Bauer:2002sh,delphifit,Mahmood:2002tt,Aubert:2004aw}.  The
idea is that using the operator product expansion (OPE), sufficiently inclusive
observables can be predicted in terms of $|V_{cb}|$, the $b$ quark mass, $m_b$,
and a few nonperturbative matrix elements that enter at order $\lqcd^2/m_b^2$
and higher orders.  One then extracts these parameters and $|V_{cb}|$ from
shapes of $B$ decay spectra and the semileptonic $B$ decay rate.  This program
also tests the consistency of the theory and the accuracy of the theoretical
predictions for inclusive decay rates. This is important also for the
determination of $|V_{ub}|$, whose error is a major uncertainty in the overall
constraints on the unitarity triangle.

The OPE shows that in the $m_b \gg \lqcd$ limit  inclusive $B$ decay rates are
equal to the $b$ quark decay rates~\cite{OPE,book}, and the corrections are
suppressed by powers of $\alpha_s$ and $\lqcd/m_b$. High-precision comparison of
theory and experiment requires a precise determination of the heavy quark
masses, as well as the nonperturbative matrix elements that enter the
expansion. These are $\lambda_{1,2}$, which parameterize the nonperturbative
corrections to inclusive observables at ${\cal O}(\lqcd^2/m_b^2)$.  At order
$\lqcd^3/m_b^3$, six new matrix elements occur, usually denoted by $\r{1,2}$ and
$\t{1,2,3,4}$. 

In this paper, we perform a global fit to the available inclusive decay
observables from BABAR, BELLE, CDF, CLEO, and DELPHI, including theoretical
expressions computed to order $\alpha_s^2 \beta_0$, $\alpha_s \lqcd/m_b$ and
$\lqcd^3/m_b^3$. A potential source of uncertainty in the OPE predictions is the
size of possible violations of quark-hadron duality~\cite{NIdual}.  Studying $B$
decay distributions is the best way to constrain these effects experimentally,
since it should influence the relationship between shape variables of different
spectra. We find that at the current level of precision, there is excellent
agreement between theory and experiment, with no evidence for violations of
duality in inclusive $b \to c$ decays.

A previous analysis of the experimental data was presented in
2002~\cite{Bauer:2002sh}. There has been considerable new data since then, which
has been included in the present analysis, and reduces the errors
on $|V_{cb}|$ and $m_b$.  In addition, the $\alpha_s \lqcd/m_b$ corrections to
the hadronic invariant mass spectrum as a function of the lepton energy cut have
now been computed~\cite{Trott:2004xc}, and are included in the present analysis.
This reduces the theoretical uncertainty on the hadronic mass moments. We also
compare our results with other recent
analyses~\cite{Gambino:2004qm,delphifit,Aubert:2004aw}.

\section{Possible Schemes}
\label{sec:schemes}

The inclusive $B$ decay spectra depend on the masses of the $b$ and $c$ quarks,
which can be treated in many different ways. The $b$ quark is treated as heavy,
and theoretical computations for $B^{(*)}$ decays are done as an expansion in
powers of $\lqcd/m_b$. The use of the $1/m_b$ expansion is common to all
methods.

The decay rates for $B \to X_c$ decay depend on the mass of the $c$ quark, for
example, through its effect on the decay phase space. One can treat the $c$
quark as a heavy quark. This allows one to compute the $D^{(*)}$ meson masses as
an expansion in powers of $\lqcd/m_c$. The observed $D^{(*)}$ masses can be used
to determine $m_c$. Since the computations are performed to $\lqcd^3/m_c^3$,
this introduces errors of fractional order $\lqcd^4/m_c^4$ in $m_c$, which gives
fractional errors of order $\lqcd^4/(m_b^2 m_c^2)$ in the inclusive $B$ decay
rates, since charm mass effects first enter at order $m_c^2/m_b^2$. In this
method, one starts with the parameters $|V_{cb}|$, $m_b$, $m_c$,
$\lambda_{1,2}$, $\r{1,2}$ and $\t{1-4}$. The $B^*-B$, $D^*-D$ and $B-D$ mass
differences can be used to eliminate $m_b-m_c$, $\lambda_2$ and $\rho_2$. Only
mass differences are used to avoid introducing the parameter $\bar \Lambda$ of
order $\lqcd$; thus we do not use the $B$ meson mass to eliminate $m_b$. Three
linear combinations of the four $\t{i}$'s occur in inclusive $B$ decays, and the
remaining linear combination would be needed for inclusive $B^*$ decays. In
summary the parameters used are  (i) $|V_{cb}|$; (ii) one parameter of order the
quark mass: $m_b$; (iii) one parameter of order $\lqcd^2$: $\lambda_1$; and (iv)
four parameters of order $\lqcd^3$: $\r1$, $\t1-3\t4$, $\t2+\t4$, $\t3+3\t4$.
These seven parameters are determined by a global fit to moments of the $B$
decay distributions, and the semileptonic branching ratio. This is the procedure
used in Ref.~\cite{Bauer:2002sh}.

An alternative approach is to avoid using the $1/m_c$ expansion for the charm
quark~\cite{Gambino:2004qm}, since it introduces $\lqcd/m_c$ corrections, which
are larger than the $\lqcd/m_b$ corrections of the $1/m_b$ expansion. In this
case heavy quark effective theory (HQET) can no longer be used for the $c$ quark
system, and there are no constraints on $m_c$ from the $D$ and $D^*$ meson
masses. At the same time, it is not necessary to expand heavy meson states in an
expansion in $1/m_{b,c}$, so that the time-ordered products $\t{1-4}$ can be
dropped. With this procedure, one has in addition to (i) $|V_{cb}|$; (ii) two
parameters of order the quark mass: $m_{b,c}$; (iii) two parameters of order
$\lqcd^2$: $\lambda_{1,2}$; and (iv) two parameters of order $\lqcd^3$:
$\r{1,2}$. The number of parameters is the same whether or not one expands in
$1/m_c$. If one does not expand, two parameters of order $\lqcd^3$ are replaced
by two lower order parameters, one of order the quark mass, and one of order
$\lqcd^2$. The expansion parameters, such as $\lambda_{1,2}$ are not the same in
the two approaches. The values of $\lambda_{1,2}$ not expanding the states in
$1/m_Q$ are the values of $\lambda_{1,2}$ plus various time-ordered products
$\t1-\t4$ when one expands the states in powers of $1/m_Q$.

In addition to the choice of expanding or not expanding in $1/m_c$, one also has
a choice of possible quark mass schemes.  It has long been known that a
``threshold mass" definition for $m_b$ is preferred over both the pole and the
\MSbar\ schemes, and it was shown in Ref.~\cite{Bauer:2002sh} that the
expansions are indeed better behaved in the $1S$~\cite{ups1,ups2} or
PS~\cite{3loopPS} schemes for $m_b$. If one expands in $1/m_c$, then $m_c$ is
eliminated through use of the meson masses, and does not enter the final
results. If one does not expand in $1/m_c$, then $m_c$ is a fit parameter. In
this method, $m_c$ is treated as much lighter than $m_b$, so the charm quark
mass is chosen to be $\mcms(\mu)$, the \MSbar\ mass renormalized at a scale $\mu
\sim m_b$. This is similar to how strange quark mass effects could be included
in $B \to X_s \gamma$ decay. In our computation, we will choose the scale
$\mu=m_b$.

In addition to the $1S$, PS, pole and \MSbar\ schemes, we have also used the
kinetic scheme mass for the $b$-quark, $\mbkin(\mu)$, renormalized at a low
scale $\mu \sim 1\,$GeV. The scale $\mu$ enters the definition of the kinetic
mass, and should not be confused with the scale parameter in dimensional
regularization.  The relation between the pole and kinetic masses is computed as
a perturbative expansion in powers of $\alpha_s(\mu)$, so one cannot make $\mu$
too small. In the kinetic scheme~\cite{Gambino:2004qm} the definitions
$\mu_\pi^2 = -\l1 + {\cal O}(\alpha_s)$, $\mu_G^2 = 3\l2$, $\rho_D^3 = \r1 +
{\cal O}(\alpha_s)$, and $\rho_{\rm LS}^3 = 3\r2$ are used.

One cannot decide which scheme is best by counting parameters, or by assuming
that not expanding in $1/m_c$ is better than expanding in $1/m_c$. Ultimately,
what matters is the accuracy to which experimentally measured quantities can be
reliably computed with currently available techniques. For example, full QCD has
two parameters, $m_b$ and $m_c$, which can be fixed using the $B$ and $D$ meson
masses. [Unfortunately, this is not possible to less than 1\% precision at the
present time.] Then one can predict all inclusive $B$ decays, as well as the
$B^*$ and $D^*$ masses with no parameters. This would be the ``best'' method to
use --- unfortunately, we cannot accurately compute the desired quantities
reliably in QCD. At present, it is better to use  the HQET expansion in $1/m_b$
and $1/m_c$, with 6 parameters, and compute to order $1/m_Q^3$. In the (distant)
future, it could well be that using full QCD, with no parameters, is the best
method to use.

We have done a fit to the experimental data using 11 schemes: the $1S$, PS,
pole, \MSbar\ and kinetic schemes expanding in $1/m_c$, not expanding in $1/m_c$
and using $\mcms(m_b)$, and finally, not expanding in $1/m_c$ and using the
kinetic scheme for both $m_b$ and $m_c$. In addition, the PS and kinetic schemes
introduce a scale $\mu$, which is sometimes called the factorization scale. We
have also examined the factorization scale dependence which is present in these
two schemes. We confirm the conclusions of Ref.~\cite{Bauer:2002sh}, that the
pole and \MSbar\ schemes are significantly worse than the threshold mass
schemes, as expected theoretically. This holds regardless of whether or not one
expands in $1/m_c$. We recommend that these schemes be avoided for high
precision fits to inclusive $B$ decays. We also find that the PS scheme gives
results comparable to those of the $1S$ scheme (both expanding and not expanding
in $1/m_c$), and that the PS scheme results do not significantly depend on the
choice of factorization scale. We compared the PS scheme with the 1S scheme in 
Ref.~\cite{Bauer:2002sh}, and do not repeat the results here.

Based on the above discussion, we present our results in five different mass
schemes, using:
\beqa\label{schemedef}
\begin{tabular}{lp{210pt}}
1.\, & \mbups\ and expand $m_b-m_c$ in terms of HQET matrix elements 
  [Scheme \oneSA], \\
2. & \mbups\ and $\mcms(m_b)$ and no time ordered products [Scheme \oneSB], \\
3. & $\mbkin(\mu=1\,\GeV)$ and expand $m_b-m_c$ [Scheme \kinA], \\
4. & $\mbkin(\mu=1\,\GeV)$ and $\mcms(m_b)$ [Scheme \kinB], \\
5. & $\mbkin(\mu=1\,\GeV)$ and $\mckin(\mu=1\,\GeV)$ [Scheme
  \kinC]. \\
\end{tabular} \nn\\*[-14pt]
\eeqa
Schemes \oneSA\ and \kinA\ contain time ordered products at order
$\lqcd^3/m_b^3$, while they are  absent from \oneSB, \kinB, and \kinC.  As
discussed, the latter three schemes have the charm quark  mass as an additional
parameter at leading order in $\lqcd/m_b$.  Scheme \oneSA\ is that used in
Ref.~\cite{Bauer:2002sh}, while scheme \kinC\ is that used in
Ref.~\cite{Gambino:2004qm}.

\section{Shape variables and the data}\label{sec:shape}

We study three different distributions, the charged lepton energy
spectrum~\cite{volo,gremmetal,GK,GS} and the hadronic invariant mass
spectrum~\cite{FLSmass1,FLSmass2,GK,Trott:2004xc} in semileptonic $B\to
X_c\ell\bar\nu$ decays, and the photon spectrum in $B\to
X_s\gamma$~\cite{FLS,kl,llmw,bauer}.  The theoretical predictions for these (as
well as for the semileptonic $B\to X_c\ell\bar\nu$ rate~\cite{LSW}) are known to
order $\alpha_s^2 \beta_0$ and $\lqcd^3 / m_b^3$, where $\beta_0=11-2n_f/3$ is
the coefficient of the first term in the QCD $\beta$-function.  For the $B\to
X\ell\bar\nu$ branching rate, we use the average of the $B^\pm$ and $B^0$ data
as quoted in the PDG~\cite{pdg},\footnote{It would be inconsistent to use the
average $b$ hadron semileptonic rate (including $B_s$ and $\Lambda_b$ states),
since hadronic matrix elements have different values in the $B/B^*$ system, and
in the $B_s/B_s^*$ or $\Lambda_b$.}
\begin{eqnarray}
{\cal B}(B\to X\ell\bar\nu) &=& 10.73 \pm 0.28\, \%.
\label{pdgbr}
\end{eqnarray}
We apply a relative $-2\%$ correction to ${\cal B}(B\to X\ell\bar\nu)$ to
account for the $B\to X_u\ell\bar\nu$ fraction, and so use
\begin{eqnarray}
{\cal B}(B\to X_c \ell\bar\nu) &=&  0.98\ {\cal B}(B\to X\ell\bar\nu) .
\label{cont}
\end{eqnarray}
The uncertainty of $|V_{ub}|$ is not a dominant error in ${\cal B}(B\to X_c
\ell\bar\nu)$. The fit result for $|V_{cb}|$ depends not only on ${\cal B}(B\to
X_c \ell\bar\nu)$, but also on the partial semileptonic branching ratios
measured by the BABAR Collaboration~\cite{Aubert:2004td}, which have smaller
errors than Eq.~(\ref{pdgbr}). The published BABAR results have already been
corrected for $B \to X_u \ell\bar\nu$ contamination.

\subsection{Lifetime}

The value of $|V_{cb}|$ depends on the $B$ meson lifetimes. The ratio of $B^+$
and $B^0$ lifetimes is $\tau_+/\tau_0=1.086 \pm 0.017$~\cite{pdg}. Isospin
violation in the $B$ meson semileptonic width is expected to be smaller than
both $\tau_+/\tau_0$ and the uncertainties in the current analysis. The $\approx
8$\% isospin violation in the lifetimes is probably due to the nonleptonic
decay channels. 

An additional source of isospin violation in the experimental measurements is
through the production rates of $B^+$ and $B^0$ mesons, which is expected to be
of the order of a few percent to perhaps as large as 10\%~\cite{kaiser}. Let
$f_+$ and $f_0$ be the fraction of $B^+$ and $B^0$ mesons produced in
$\Upsilon(4S)$ decay, with $f_++f_0=1$. Then the measured semileptonic branching
ratios are
\begin{eqnarray}
{\cal B} &=& f_+ {\Gamma_{\text{sl}} \over \Gamma_{\text{sl}} 
  + \Gamma_{\text{nl},+}}
  + f_0 {\Gamma_{\text{sl}} \over \Gamma_{\text{sl}} + \Gamma_{\text{nl},0}}
\label{8}
\end{eqnarray}
where ${\cal B}$ and $\Gamma_{\text{sl}}$ are computed with the same lepton
energy cut. In writing Eq.~(\ref{8}), we have used the fact that isospin
violation in the semileptonic rates is small, so that the same value for 
$\Gamma_{\text{sl}}$ is used for both $B^+$ and $B^0$.

The measured semileptonic branching ratios can thus be written as
\begin{eqnarray}
{\cal B} &=& \tau_{\text{eff}}\ \Gamma_{\text{sl}}
\end{eqnarray}
in terms of the effective lifetime
\begin{eqnarray}
\tau_{\text{eff}} &=& f_+ \tau_+ + f_0 \tau_0.
\label{9a}
\end{eqnarray}
One can rewrite this as
\begin{eqnarray}
\tau_{\text{eff}} &=& {\tau_+ + \tau_0 \over 2} 
  + {(f_+ - f_0)(\tau_+ - \tau_0) \over 2}.
\label{9b}
\end{eqnarray}
Using the PDG 2004 lifetime values, and the measured $f_+/f_0$
ratio~\cite{isospin} gives
\begin{eqnarray}
\tau_{\text{eff}} &=& 1.60 \pm 0.01\ \text{ps}
\label{9}
\end{eqnarray}
where the contribution from the second term in Eq.~(\ref{9b}) is negligible to
both the value and the error.

\subsection{Lepton Moments}

For the charged lepton energy spectrum we define the integrals
\beq\label{Rdef}
R_n(\Ecut,\mu) =  \int_{\Ecut} \left(E_\ell-\mu\right)^n\,
  {\d\Gamma\over{\rm d}E_\ell}\, \d E_\ell\,,
\eeq
where $\d\Gamma/\d E_\ell$ is the spectrum in the $B$ rest frame and $\Ecut$ is
a lower cut on the lepton energy.  Moments of the lepton energy spectrum with a
lepton energy cut $\Ecut$ are given by
\beq
\vev{E_\ell^n}_{\Ecut} = {R_n(\Ecut,0) \over 
  R_0(\Ecut,0)}\,,
\eeq
and central moments by
\beq
\vev{\left(E_\ell-\vev{E_\ell}\right)^n}_{\Ecut} 
  = {R_n(\Ecut,\vev{E_\ell}_{\Ecut}) \over 
  R_0(\Ecut,0)}\,,
\eeq
which can be determined as a linear combination of the non-central moments.

The BABAR Collaboration~\cite{Aubert:2004td} measured the partial branching
fraction $\tau_B R_0(\Ecut,0)$, the mean lepton energy
$\vev{E_\ell}_{\Ecut}$, and the second and third central moments
$\vev{\left(E_\ell-\vev{E_\ell}\right)^n}_{\Ecut}$ for $n=2,3$, each
for lepton energy cuts of $\Ecut=0.6,\ 0.8,\ 1.0,\ 1.2$, and $1.5$~GeV.

The BELLE Collaboration~\cite{Abe:2004zv} measured the mean lepton energy and the second central moment for lepton energy cuts of $\Ecut=0.6,\ 0.8,\ 1.0,\ 1.2$, and $1.5$~GeV.

The CLEO Collaboration~\cite{Mahmood:2002tt,Mahmood:2004kq} measured the mean
lepton energy and second central moment (variance) for $\Ecut=0.7-1.6$~GeV in
steps of $0.1$~GeV.

The DELPHI Collaboration~\cite{DELPHIdata} measured the mean lepton energy, and
the $n=2,3$ central moments, all with no energy cut.

In total, we have 53 experimental quantities from the lepton moments, 20 from
BABAR, 10 from BELLE, 20 from CLEO, and 3 from DELPHI.

\subsection{Hadron Moments}
\label{subsec:hadron}

For the $B\to X_c\ell\bar\nu$ hadronic invariant mass spectrum, we define
\beq\label{Sdef}
\mX{2n}_{\Ecut} = {\displaystyle \int_{\Ecut} (m_X^2)^n\,
  {\d\Gamma\over{\rm d}m_X^2}\, \d m_X^2 \over \displaystyle
  \int_{\Ecut} {\d\Gamma\over{\rm d}m_X^2}\, {\rm d}m_X^2 }\,,
\eeq
where $\Ecut$ is again the cut on the lepton energy.  Sometimes
$\mDbar^2 \equiv [(m_D+3m_{D^{*}})/4]^2$ is subtracted out in the definitions,
$\langle (m_X^2-\mDbar^2)^n\rangle$, or the measurements of the normal moments
are quoted, $\langle (m_X^2-\langle m_X^2\rangle)^n\rangle$, but these can
easily be computed from \mX{2n}.

The BABAR Collaboration~\cite{Aubert:2004te} measured the mean values of
$m_X$, $m_X^2$, $m_X^3$ and $m_X^4$ (i.e., $n=1/2,\ 1,\ 3/2,\ 2$ moments) for
lepton energy cuts $\Ecut=0.9-1.6$~GeV in steps of $0.1$~GeV.

The BELLE Collaboration~\cite{Abe:2004ks} measured the mean values of $m_X$ and $m_X^2$ for lepton energy cuts $\Ecut=0.9-1.6$~GeV in steps of $0.1$~GeV.

The CDF Collaboration~\cite{cdfhadron} measured the mean value of $m_X^2$ and
its variance, with a lepton energy cut $\Ecut=0.7$~GeV.

The CLEO Collaboration~\cite{Csorna:2004kp} measured the mean value of 
$m_X^2-\overline m_D^2$ and the variance of $m_X^2$ for lepton energy cuts of 
$1.0$ and $1.5$~GeV.

The DELPHI Collaboration~\cite{DELPHIdata} measured the mean value of
$m_X^2-\overline m_D^2$, $(m_X^2-\overline m_D^2)^2$, the variance of $m_X^2$, and
the third central moment of $m_X^2$, all with no energy cut.

Recently half-integer moments of the $m_X^2$
spectrum~\cite{Gambino:2004qm,Trott:2004xc} have received some attention.  While
non-integer moments of the lepton energy spectrum have been computed in a power
series in $1/m_b$~\cite{BT}, this is not true for fractional moments of the
$m_X^2$ spectrum. In~\cite{Gambino:2004qm,Trott:2004xc} expressions for the
half-integer moments were proposed which involve expansions that were claimed to
be in powers of $\lqcd/m_c$. However, in the limit  $m_c \ll m_b$ (i.e., $m_c$
of order $\sqrt{m_b\lqcd}$ or less), the higher order terms in these expansion
scale with powers of $m_b\lqcd/m_c^2$, which in this limit is of order unity or
larger.  On the other hand, in the small velocity limit, $m_b \sim m_c \gg
m_b-m_c \gg \lqcd$, the expansion of \mX{2n+1} is well-behaved. Thus, the
calculations of the half-integer moments as presented
in~\cite{Gambino:2004qm,Trott:2004xc} do not correspond to a power series in
$1/m_{b,c}$ in the $m_c \ll m_b$ limit and omitted terms are only power
suppressed in the small velocity limit. In addition, the BLM corrections to
these moments are currently unknown, because they require the BLM contribution
from the virtual terms, which have not been computed. For these reasons we will
not use these half-integer moments in the fit, but will compare the fit results
with the measured values. Omitting the half-integer moments, there are 16 data
points from BABAR, 8 from BELLE, 2 from CDF, 4 from CLEO, and 4 from DELPHI, for a total of 34 measurements.

\subsection{Photon Spectrum}

For $B\to X_s\gamma$, we define 
\beq\label{Tdef}
\Eg{n}_{\Ecut} =  {{\displaystyle \int_{\Ecut} E_\gamma^n\,
  {\d\Gamma\over \d E_\gamma}\, \d E_\gamma} \over {\displaystyle 
  \int_{\Ecut} {\d\Gamma\over \d E_\gamma}\, \d E_\gamma}} \,,
\eeq
where $\d\Gamma/\d E_\gamma$ is the photon spectrum in the $B$ rest frame, and
$\Ecut$ is the photon energy cut.  In this case the variance, $\langle (E_\gamma
- \Eg{})^2 \rangle = \Eg2-\Eg{}^2$, is often used instead of the second moment,
and higher moments are not used as they are very sensitive to the boost of the
$B$ meson in the $\Upsilon$ rest frame (though this is absent if $\d \Gamma/\d
E_\gamma$ is reconstructed from a measurement of $\d \Gamma/\d E_{m_{X_s}}$) and
to the details of the shape function.   \Eg{n} are known to order $\alpha_s^2
\beta_0$~\cite{llmw} and $\lqcd^3 / m_b^3$~\cite{bauer}.  These moments are
expected to be described by the OPE  once $m_B/2-\Ecut \gg \lqcd$.  Precisely
how low $\Ecut$ has to be to trust the results can only be decided by studying
the data as a function of $\Ecut$; one may expect that $\Ecut = 1.8\,\GeV$
available at present~\cite{Koppenburg:2004fz} is sufficient.  Note that the
perturbative corrections included are sensitive to the $m_c$-dependence of the
$b\to c\bar c s$ four-quark operator ($O_2$) contribution.  This is a
particularly large effect in the $O_2-O_7$ interference~\cite{llmw}, but its
relative influence on the moments of the spectrum is less severe than that on
the total decay rate.  

We use the BELLE~\cite{Koppenburg:2004fz}, and CLEO~\cite{cleophoton}
measurements of the mean photon energy and variance, with photon energy cuts of
$1.8$ and $2.0$~GeV, respectively, and the BABAR
measurement~\cite{Aubert:2002pb} of the mean photon energy with a cut of
$2.094$~GeV for a total of 5 measurements.

\section{Fit Procedure}
\label{sec:mass}

As discussed in Sec.~\ref{sec:schemes}, there are many ways to treat the quark
masses and hadronic matrix elements that occur in the OPE results for the
spectra. In the schemes where $m_b-m_c$ is expanded in HQET (such as \oneSA\ and
\kinA), the theoretical expressions for the shape variables defined in
Eqs.~(\ref{Rdef}), (\ref{Sdef}), and (\ref{Tdef}) include 17 terms
\beqa\label{expdef}
X_{\Ecut} &=& X^{(1)} + X^{(2)}\, \Lambda 
  + X^{(3)}\, \Lambda^2 + X^{(4)}\, \Lambda^3
  + X^{(5)}\, \l1 \nn\\*
  &+&  X^{(6)}\, \l2 
  + X^{(7)}\, \l1 \Lambda + X^{(8)}\, \l2 \Lambda + X^{(9)}\, \r1  \nn\\*
&+& X^{(10)}\, \r2 
  + X^{(11)}\, \t1 + X^{(12)}\, \t2
  + X^{(13)}\, \t3  \nn\\*
&+&  X^{(14)}\, \t4 + X^{(15)}\, \epsilon + X^{(16)}\, \epsBLM
  + X^{(17)}\, \epsilon\, \Lambda\,, \nn\\*
\eeqa
while in the schemes when $m_c$ is treated as an independent free parameter
(such as \oneSB, \kinB, and \kinC), we have 22 terms
\beqa\label{expdef2}
Y_{\Ecut} &=& Y^{(1)} + Y^{(2)}\, \Lambda + Y^{(3)}\, \Lambda_c
  + Y^{(4)}\, \Lambda^2 + Y^{(5)}\, \Lambda \Lambda_c  \nn\\*
&+&  Y^{(6)}\, \Lambda_c^2 + Y^{(7)}\, \Lambda^3 + Y^{(8)}\, \Lambda^2 \Lambda_c
  + Y^{(9)}\, \Lambda \Lambda_c^2  \nn\\*
  &+&  Y^{(10)}\, \Lambda_c^3
  + Y^{(11)}\, \l1 + Y^{(12)}\, \l2 + Y^{(13)}\, \l1 \Lambda  \nn\\*
&+&  Y^{(14)}\, \l2 \Lambda
  + Y^{(15)}\, \l1 \Lambda_c + Y^{(16)}\, \l2 \Lambda_c
  + Y^{(17)}\, \r1 \nn\\*
  &+& Y^{(18)}\, \r2 + Y^{(19)}\, \epsilon + Y^{(20)}\, \epsBLM
  + Y^{(21)}\, \epsilon\, \Lambda \nn\\*
  &+& Y^{(22)}\, \epsilon\, \Lambda_c \,.
\eeqa
In Eqs.~(\ref{expdef}) and (\ref{expdef2}) $\Lambda$ and $\Lambda_c$ are
respectively the differences between the $b$ and $c$ quark masses and their
reference values about which we expand.  The coefficients $X^{(k)}$ and
$Y^{(k)}$ are functions of $\Ecut$, and ($X_{\Ecut}$, $Y_{\Ecut}$) are any of
the experimental observables discussed earlier. The parameter $\epsilon \equiv
1$ counts powers of $\alpha_s$. We have used $\alpha_s(m_b) = 0.22$.  The strong
coupling constant is not a free parameter, but is determined from other
measurements such as the hadronic width of the $Z$. The hadron and lepton
moments are integrals of the same triple differential decay rate with different
weighting factors. The use of different values of $\alpha_s$ for the hadron and
lepton moments, as done in Ref.~\cite{Gambino:2004qm}, is an ad hoc choice.

\section{The Fit}
\label{sec:fit}

We use the program MINUIT to perform a global fit to all observables introduced
in Sec.~\ref{sec:shape} in each of the 11 schemes mentioned in
Sec.~\ref{sec:schemes}.  There are a total of 92 lepton, hadron, and photon
moments, plus the semileptonic width, to be fit using 7 parameters, so the fit
has $\nu=86$ degrees of freedom.

To evaluate the $\chi^2$ required for the fit, we include both experimental and
theoretical uncertainties. For the experimental uncertainties we use the full
correlation matrix for the observables from a given differential spectrum as
published by the experimental collaborations.  In addition to these experimental
uncertainties there are theoretical uncertainties, which correspond to how well
we expect to be able to compute each observable theoretically.  For a given
observable, our treatment of theoretical uncertainties is similar to that in
Ref.~\cite{Bauer:2002sh}. 

It is important to include theoretical uncertainties in the fit, since not all
quantities can be computed with the same precision. We have treated theoretical
errors as though they have a normal distribution with zero mean, and standard
deviation equal to the error estimate.\footnote{This is the same procedure as
that used in doing a fit to the fundamental constants~\cite{CODATA}.
An example which makes clear why theoretical errors should be included is: The
Hydrogen hyperfine splitting is measured to 14 digits, but has only been
computed to 7 digits. The Positronium hyperfine splitting is measured and
computed to 8 digits. It would not be proper to give the H hyperfine splitting a
weight $10^6$ larger than the Ps hyperfine splitting in a global fit to the
fundamental constants.} Strictly speaking, the theoretical formula has some
definite higher order correction, which is at present unknown. One can then view
the normal distribution used for the theory value as the prior distribution in a
Bayesian analysis. The way in which theoretical errors are included is a matter
of choice, and there is no unique prescription.

We now discuss in detail the theoretical uncertainties included in the fit.
Those who find this procedure abhorrent can skip the entire discussion, since we
will also present results not including theory errors.

\subsection{Theory uncertainties}

Theoretical uncertainties in inclusive observables as discussed here originate
from four main sources.  First, there are uncertainties due to uncalculated
power corrections.  For schemes \oneSA\ and \kinA, these are of order
$\lqcd^4/(m_b^2 m_c^2) \sim 0.001$, while for schemes \oneSB\ and \kinB\ where
no $1/m_c$ expansion is performed, these are of order $\lqcd^4/m_b^4 \sim
0.0001$.  Next, there are uncertainties due to uncalculated higher order
perturbative terms.  In particular, the full two loop result proportional to
$\alpha_s^2/(4\pi)^2 \sim 0.0003$ is not available. An alternative way to
estimate these perturbative uncertainties is by the size of the smallest term
computed in the series, which is the term proportional to $\alpha_s^2 \beta_0$. 
We choose here to use half of this last computed term as an estimate of the
uncertainty.  There are also uncalculated effects of order $(\alpha_s/4 \pi)
\lqcd^2/m_b^2 \sim 0.0002$.  Finally, there is an uncertainty originating from
effects not included in the OPE in the first place.  Such effects sometimes go
under the name ``duality violation," and are very hard to quantify. For this
reason, we do not include an explicit contribution to the overall theoretical
uncertainty from such effects.  If duality violation would be larger than the
other theoretical uncertainties they would give rise to a poor fit to the data.
To determine the uncertainties for dimensionful quantities such as the moments
considered here, we have to multiply these numbers by the appropriate
dimensionful quantity. This number is obtained from dimensional analysis, and we
use for the $n$'th hadronic moment $(m_B^2)^n f_n$, while we use $(m_B/2)^n f_n$
for the $n$'th leptonic moment. The factors $f_n$ are chosen to be $f_0=f_1=1$,
$f_2=1/4$ and $f_3=1/(6 \sqrt{3})$. The values for $f_2$ and $f_3$ are the
maximum allowed values for the second and third central moments (variance and
skewness) for a probability distribution on the interval $[0,1]$.

The complete BLM piece has not been computed for the non-integer hadronic
moments. The perturbative uncertainty is therefore dominated by this
contribution of order $\beta_0 \alpha_s^2/(4 \pi)^2 \sim 0.003$. We will use $A
\sim 0.003$ for the non-integer hadronic moments when we compare experiment with
theory.

For the hadronic mass and lepton energy moments, which depend on the value of
the  cut on the lepton energy, we have to decide how to treat the correlation of
the theoretical  uncertainties. In the global fit by the BABAR
Collaboration~\cite{Aubert:2004aw}, the theory errors for a given observable
with different cuts on $E_\ell$ were treated as 100\% correlated.  This ignores
the fact that the  higher order terms omitted in the OPE depend on the lepton
energy cut. In Ref.~\cite{Bauer:2002sh}, only the two extreme values of the
lepton energy cut were included in the fit, and the correlation of the theory
uncertainties was neglected.  Here we take the correlation of the theoretical
uncertainties to be given by the correlation between the experimental
measurements, which captures the correlations due to the fact that observables
with different cuts share some common events.

For the photon energy moments an additional source of uncertainty is the fact
that  the presence of any experimentally sensible value for $\Ecut$ affects the
mean photon energy $\vev{E_\gamma}$ such that the extracted value of $m_b$ is
biased toward larger values because of shape function effects~\cite{bauer}. 
However, this shift cannot be calculated model independently. Rather than
include a model dependence, we have multiplied the theory uncertainties for the
$b \to s \gamma$ rates by the ratios of the energy difference from the endpoint,
relative to that for BELLE with $\Ecut=1.8$~GeV.\footnote{The $B\to X_s\gamma$
photon spectrum also receives contributions of order $m_s^2/m_b^2$, which are
negligible corrections for our analysis.}

To summarize, we define the combined experimental and theoretical error matrix
for a given observable to be
\beq
\sigma^2_{ij} = \sigma_i\, \sigma_j\, c_{ij} \,,\qquad \text{no\ sum on}\ i,j\,,
\eeq
where $i$ and $j$ denote observables, $c_{ij}$ is the experimental correlation
matrix, and 
\beqa
\sigma_i &=& \sqrt{(\sigma^{\rm exp}_i)^2 + \left( A\, f_n m_B^{2n} \right)^2 
  + \left({B_i}/{2}\right)^2}  \nn\\*
&& \mbox{for the n'th hadron moment\,,} \nn\\
\sigma_i &=& \sqrt{(\sigma^{\rm exp}_i)^2 + \left(A\, f_n {(m_B/2)^{n}}\right)^2
  + \left({B_i}/{2}\right)^2}  \nn\\*
&& \mbox{for the n'th lepton moment\,,} \nn\\
\sigma_i &=& \sqrt{(\sigma^{\rm exp}_i)^2 + \left(A\, f_n {(m_B/2)^{n}}\right)^2
  + \left({B_i}/{2}\right)^2}  \nn\\*
  && \mbox{for  the n'th photon moment\,,} \label{adef}
\eeqa
and $f_0=f_1=1$, $f_2=1/4$, $f_3=1/(6 \sqrt{3})$. Here $\sigma^{\rm exp}_i$ are
the experimental errors, $B_i=X^{(16)}$  or $Y^{(20)}$ are the coefficients of
the last computed terms in the perturbation series, and $A$ contains the errors
discussed earlier. We take $A=0.001$ for the data used in the fit, except for
the CLEO and BABAR photon moments, where we multiply $A$ by $1.3$ and $1.5$,
respectively, to account for the increase in shape function effects as one
limits the allowed region of the photon spectrum.

We stress that there is no unique way to estimate theoretical uncertainties to a
given expression.  Thus, while we believe that our estimates are reasonable, it
is certainly not the only possible way to estimate the theory uncertainties
(e.g., taking the theory correlation to be identical to the experimental
correlations is just an educated guess).

\subsection{Experimental correlations}

Some of the experimental correlation matrices have negative eigenvalues. In some
cases, these are at the level of round-off errors. To avoid these negative
values, we have added $0.01$ to the diagonal entries for the correlation
matrices for the BABAR and CLEO lepton moments, and the DELPHI hadron moments.

The correlation matrix for the BABAR hadronic moments~\cite{Aubert:2004aw}
contains negative eigenvalues which are much larger than any round-off
uncertainties. This persists even if only every second value of the cut is used,
as advocated in \cite{Aubert:2004aw}, so we are forced to add $0.05$ to the
diagonal entries of the correlation matrix for the BABAR hadron moments to make
the eigenvalues positive. Note that the correlation matrix can have negative
eigenvalues only if the probability distribution can take on negative values.

The preliminary correlation matrix for the BELLE lepton and hadron moments was
used in the fit~\cite{belleprivate}.

\begin{figure*}[tbp]
\centerline{\includegraphics[bb=20 170 580 690,height=10cm]{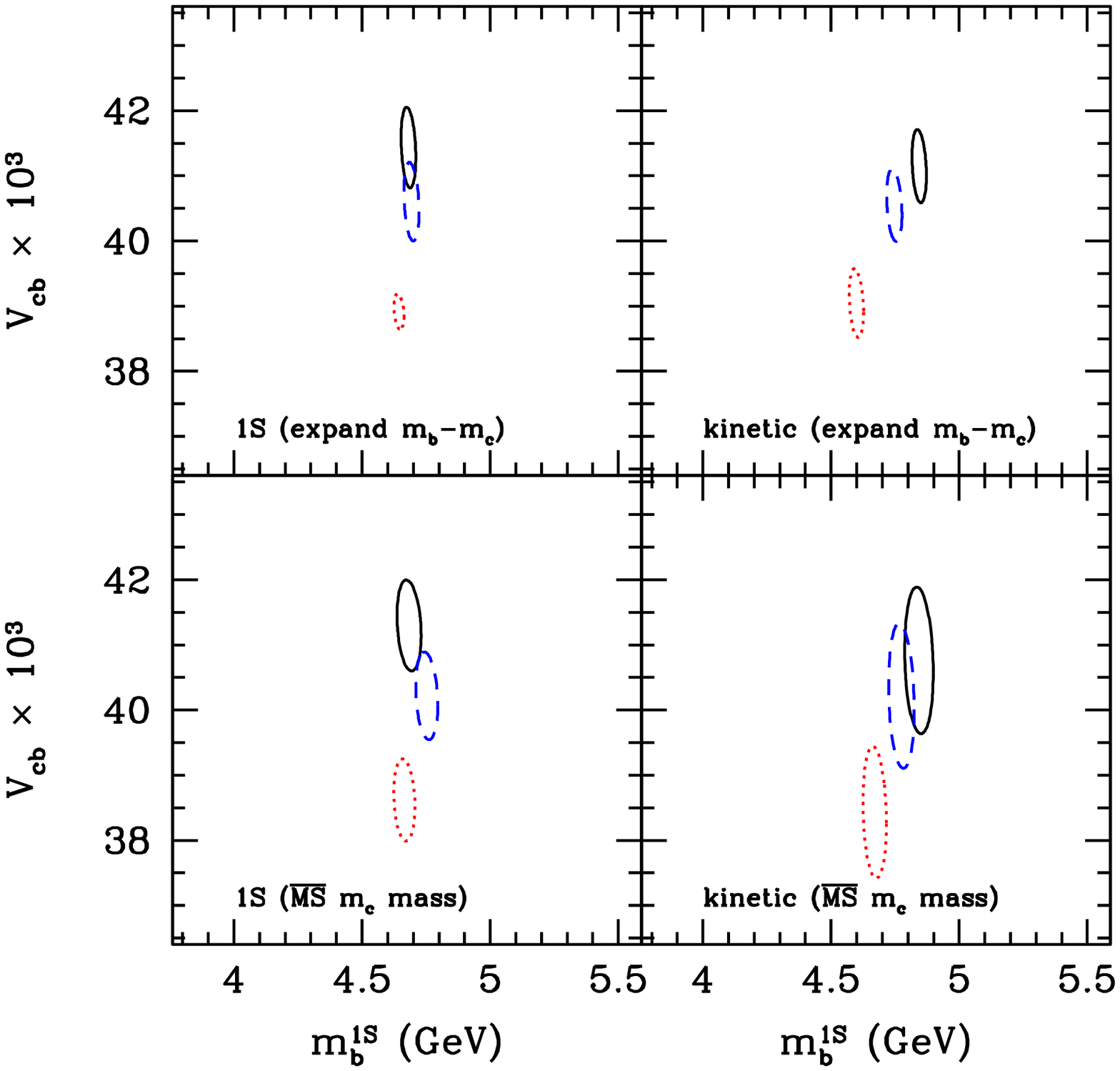}
\includegraphics[bb=30 170 350 690,height=10cm]{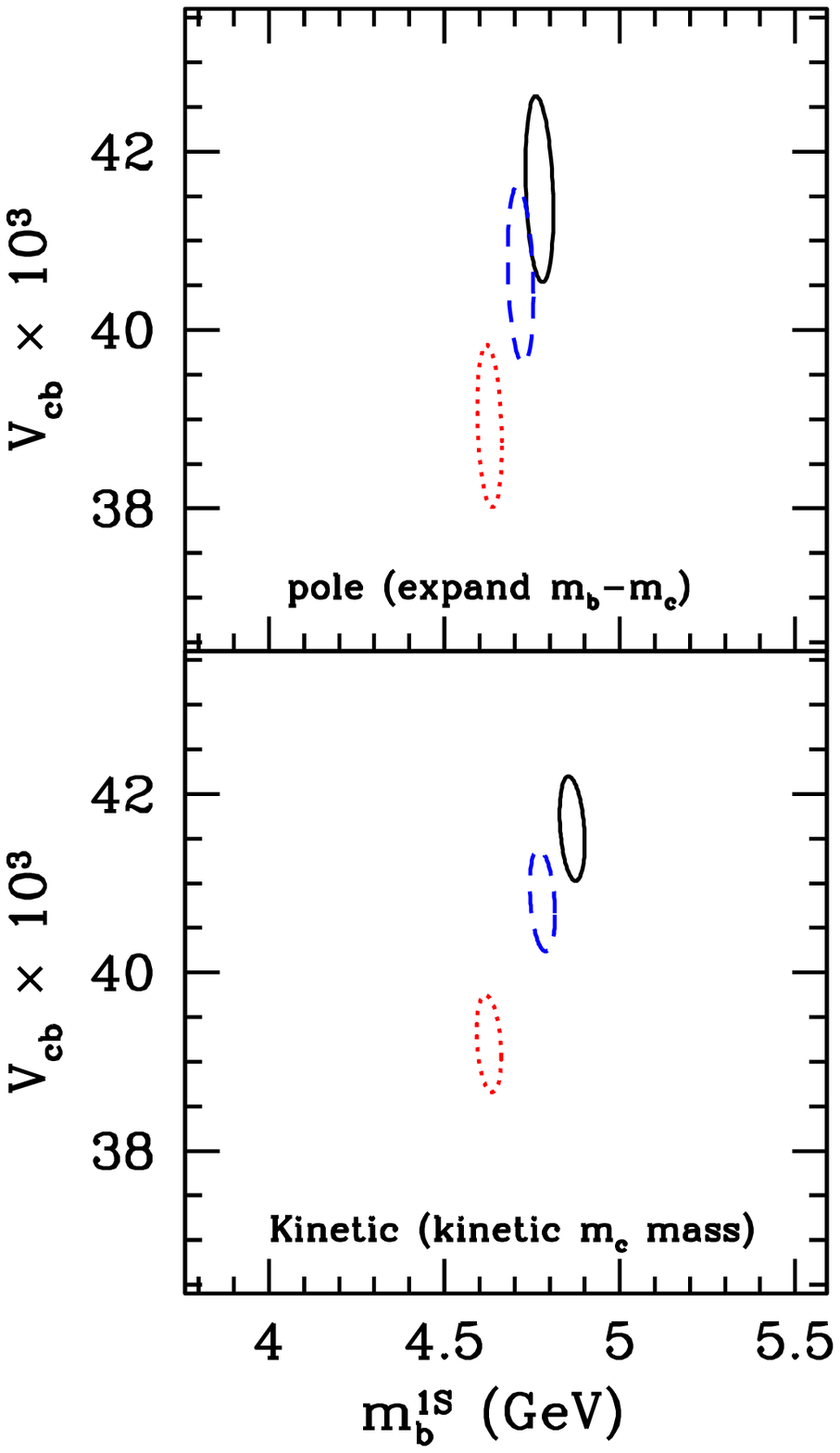}}
\caption{Fit results for $|V_{cb}|$ and $m_b$ in the \oneSA, \oneSB, \kinA, 
\kinB, and \kinC\ schemes defined in Eq.~(\ref{schemedef}) and in the
traditional pole scheme.  The red doted, blue dashed, and black solid ellipses
denote the results at tree level, order $\alpha_s$, and $\alpha_s^2\beta_0$,
respectively, corresponding to $\Delta\chi^2 = 1$.}
\label{fig:schemecompare}
\end{figure*}

\begin{table*}[tbp]
\begin{tabular}{c|c||c|c|c|c|c|c}
~Scheme~  &  ~$\sigma^2_{\text{theory}}$~  &  ~~~$\chi^2/\nu$~~~  &  
  ~$|V_{cb}|\times 10^3$~  &  ~$m_b^{1S}\, [\GeV]$~  &  ~$m_b-m_c \, [\GeV]$~  &
  ~$\overline{m}_c(\overline{m}_c)\, [\GeV]$~  &  ~~~$\lambda_1\, [\GeV^2]$~~~
\\
\hline\hline
\oneSA & yes & $50.9/86$ &
$41.4\pm 0.6$ &
$4.68\pm 0.03$ & 
$3.41 \pm 0.01$ &
$1.07 \pm 0.04$ &
$-0.27\pm 0.04$  
\\ \hline
\kinA& yes&  $52.6/86$ &
$41.2 \pm 0.6 \pm 0.1 $ &
$4.70\pm 0.03 \pm 0.03$ & 
$3.40 \pm 0.01 \pm 0.01 $ &
$1.09 \pm 0.03 \pm 0.03 $ &
$-0.19\pm 0.04 \pm 0.04 $ 
\\ \hline\hline
\oneSA& no & $148.4/86$ &
$41.5\pm 0.3$ &
$4.69\pm 0.02$ & 
$3.39 \pm 0.01$ &
$1.09 \pm 0.02$ &
$-0.31\pm 0.03$ 
\\ \hline
\kinA& no & $238.8/86$ &
$41.1 \pm 0.3 \pm 0.7$ &
$4.74\pm 0.01 \pm 0.11$ & 
$ 3.36 \pm 0.01 \pm 0.04 $ &
$1.15 \pm 0.01 \pm 0.11$ &
$-0.33\pm 0.03 \pm 0.11$ 
\\ \hline\hline
\end{tabular}
\caption{Fit results for $|V_{cb}|$, $m_b$, $m_c$ and $\lambda_1$  in the
\oneSA\  and \kinA\  schemes. Our fits in the kinetic scheme use $\mu_\pi^2$,
but the result is converted to $\lambda_1$ to help comparison. The
first two lines are the fit results including our estimates of the theoretical
errors, the lower two lines are setting these to zero.  The second error for the
\kinA\ scheme is the shift due to changing $\mu$ from 1 to 1.5~GeV.}
\label{tab:central}
\end{table*}

\subsection{Constraints on parameters}

Even though there are many more observables than there are parameters, the fit
does not provide strong constraints on the $1/m_b^3$ parameters. Thus it is
useful to add additional information to ensure that the fit converges to
physically sensible values of the nonperturbative parameters. Thus, as in
Ref.~\cite{Bauer:2002sh} we add to $\chi^2$ the contribution
\beq\label{delchi}
\chi^2_{\text{param}}(m_\chi,M_\chi) = \cases{
0\,,  &  $|\langle {\cal O} \rangle| \le m_\chi^3$\,, \cr
\left[|\langle \mathcal{O} \rangle| - m_\chi^3\right]^2 / M_\chi^6 \,,
  &  $|\langle {\cal O} \rangle| > m_\chi^3$\,, \cr}
\eeq
where $(m_\chi,M_\chi)$ are both  quantities of order $\lqcd$, and  $\langle
{\cal O} \rangle$ are the matrix elements of any of the $1/m^3$ operators in the
fit. This way we do not prejudice $\langle \mathcal{O} \rangle$ to have any
particular value in the range $|\langle {\cal O} \rangle| \le m_\chi^3$.  In the
fit we take $M_\chi=m_\chi=500\,\MeV$. We checked in Ref.~\cite{Bauer:2002sh}
that  the results for $|V_{cb}|$ and $m_b$ are insensitive to varying $m_\chi$
between 500~MeV and 1~GeV. The data are sufficient to constrain the $1/m_b^3$
operators in the sense that they can be consistently fit with reasonable values,
but they are not determined with any useful precision. The data can be fit
without including $\chi^2_{\text{param}}$, but then some of the $1/m_b^3$
parameters are not of natural size, with values of order $0.5~\text{GeV}^3$.
Including $\chi^2_{\text{param}}$ gives a fit with reasonable values of the
parameters, of order $0.1~\text{GeV}^3$. The contribution of
$\chi^2_{\text{param}}$ is rather small, of order $0.1-0.2$, so that
$\chi^2_{\text{param}}$ does not drive the fit. This shows that there are some
very flat directions in parameter space which are stabilized by including
$\chi^2_{\text{param}}$. We have shown our final results for $V_{cb}$ and $m_b$
with and without including $\chi^2_{\text{param}}$ in the fit. The final results
do not depend significantly on whether or not $\chi^2_{\text{param}}$ is
included.

Note that the fit performed by the BABAR Collaboration included the half-integer
hadronic moments. We have checked that including these moments still leaves some
$1/m_b^3$ parameters with values larger than natural size. We have chosen to not
include these moments in the fit since they have large theoretical
uncertainties.

\begin{figure*}[t]
\centerline{\includegraphics[bb=70 170 590 710,width=0.48\textwidth]{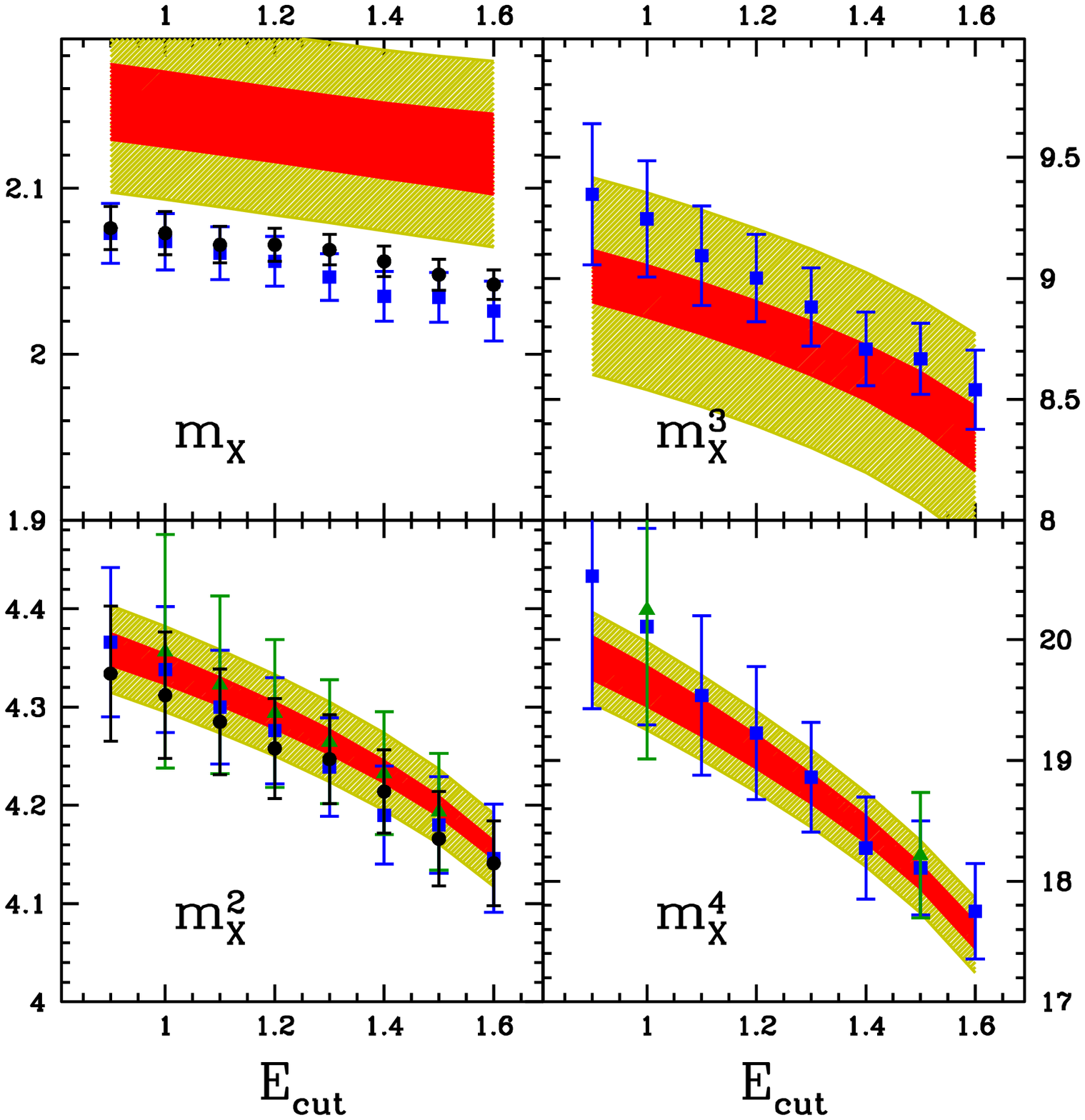}
\hfill \includegraphics[bb=70 170 590 710,width=0.48\textwidth]{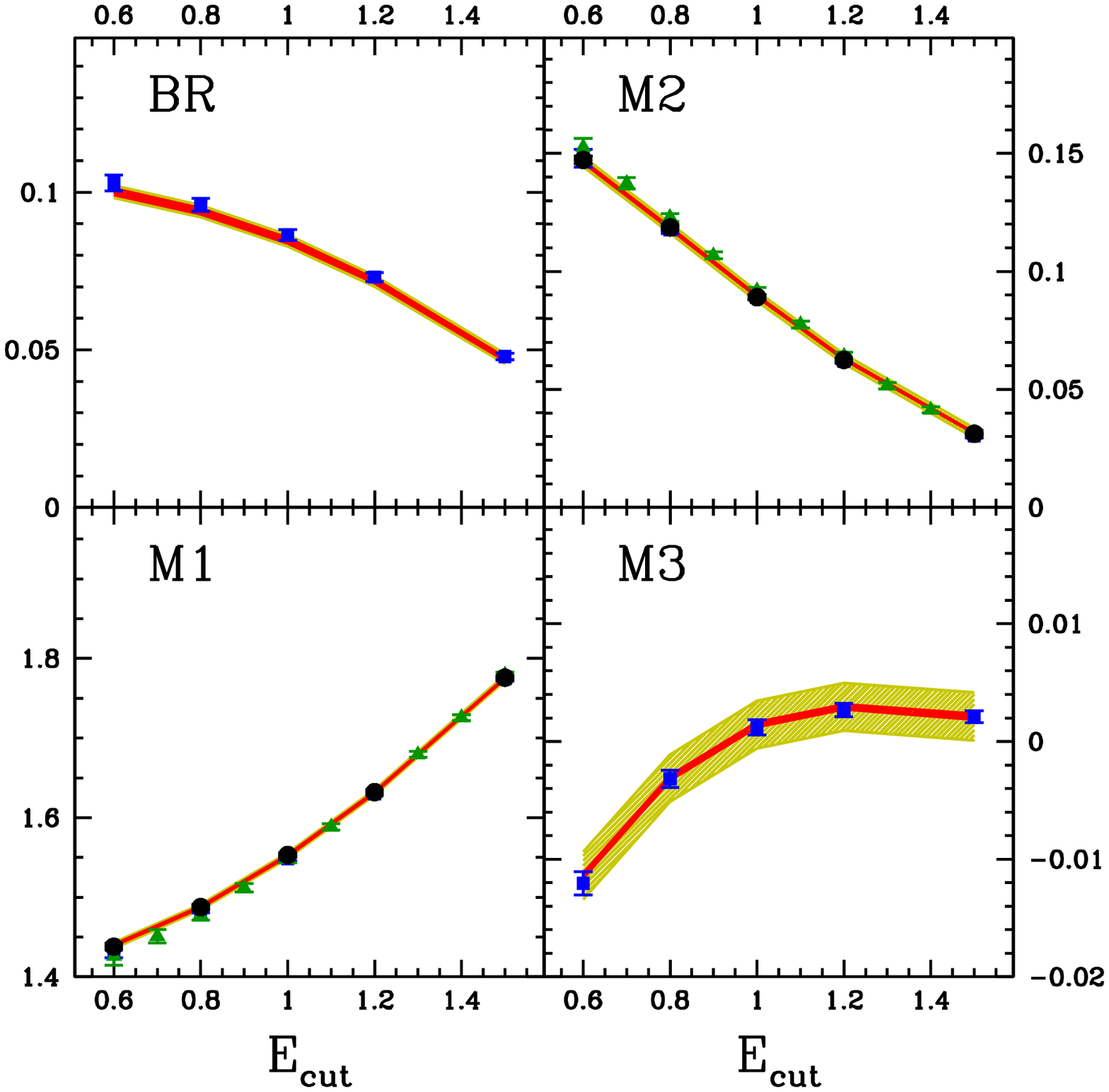}}
\caption{Measurements (blue squares:
BABAR~\protect\cite{Aubert:2004te,Aubert:2004td}, green triangles:
CLEO~\protect\cite{Csorna:2004kp,Mahmood:2004kq}, black dots:
BELLE~\protect\cite{Abe:2004ks,Abe:2004zv}) and fit results for the hadron
invariant mass (left) and the lepton energy moments (right) as functions of the
lepton energy cut, $\Ecut$.  For the hadron moments $m_X^n$ denotes \mX{n},
while for the lepton moments BR is branching ratio, M1 is first moment, and M2
and M3 are the second and third central moments, respectively.  The red (dark)
shaded regions show the fit error, while the yellow (light) shaded regions are
our estimates of the theoretical uncertainties from the $A$ terms in
Eq.~(\ref{adef}). The $A$ term for \mX{} and \mX3 is three times larger than for
\mX2 and \mX4, because of the worse expansion for the non-integer moments.}
\label{fig:hadlepcompare}
\end{figure*}

\begin{figure*}[t]
\centerline{\includegraphics[bb=70 170 590 710,width=0.48\textwidth]{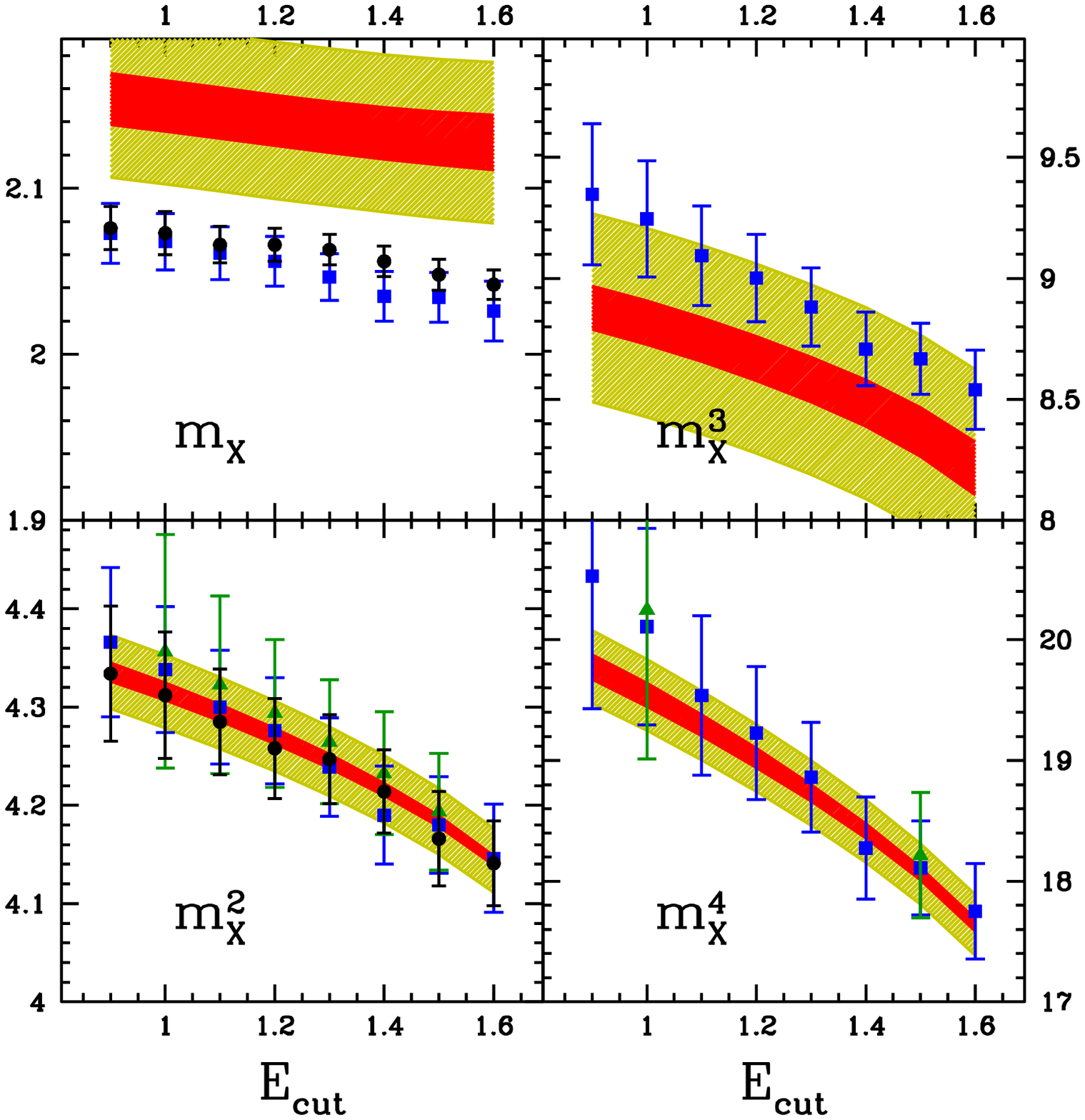}
\hfill \includegraphics[bb=70 170 590 710,width=0.48\textwidth]{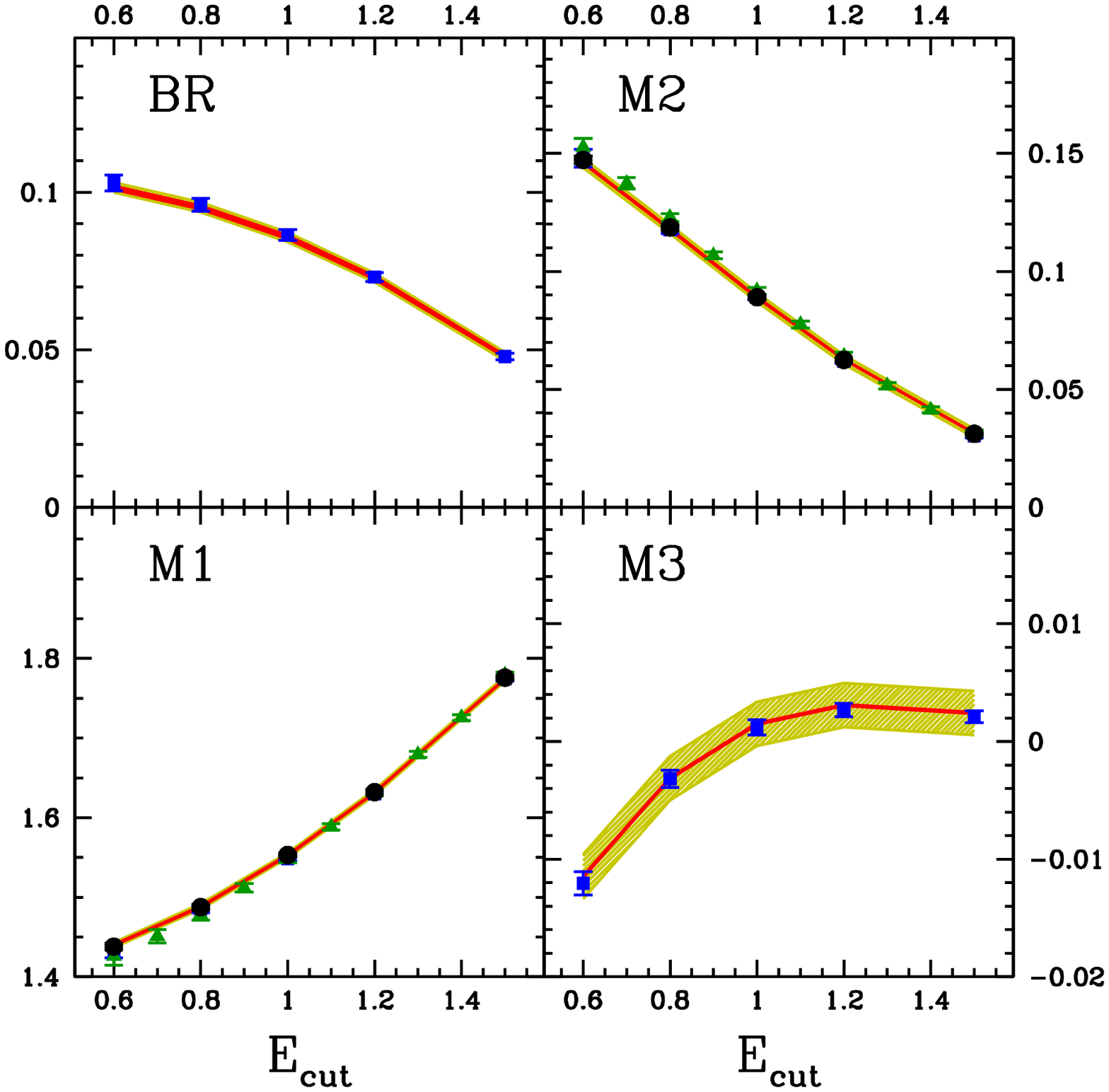}}
\caption{Measurements and fit results for the hadron invariant mass and the
lepton energy moments, setting all theory errors to zero in the fit.  (See the
caption for Fig.~\ref{fig:hadlepcompare}.) The yellow (light) shaded band gives
the estimated theoretical uncertainty, as in Fig~\ref{fig:hadlepcompare}. It is
not included in the fit, but it can help to decide the significance of 
any differences between theory and experiment.
\label{fig:nothcompare}}
\end{figure*}

\section{Fit Results and Discussion}
\label{sec:results}

The fit result for $|V_{cb}|$ and $m_b$ in the five mass schemes defined in
Eq.~(\ref{schemedef}) and in the traditional pole scheme are shown in
Fig.~\ref{fig:schemecompare}. 
The fit results are shown at tree level, order $\alpha_s$, and order $\alpha_s^2
\beta_0$. The kinetic scheme results are obtained using $\mu_\pi^2$ etc.\ in the
fit, and then converting the results back to $\lambda_1$ etc.\ for easier
comparison with the other schemes. One can see that the \oneSA\ and \kinA\
schemes have better convergence than the pole scheme.  The main fit results in
the \oneSA\ and \kinA\ scheme are given in Table~\ref{tab:central}.  The quoted
$|V_{cb}|$ values include electromagnetic radiative corrections, that
reduce\footnote{In the preprint version of this paper and in
Refs.~\cite{Bauer:2002sh,ups1} the inverse of this factor was used erroneously,
which enhanced $|V_{cb}|$.  We thank O.\ Buchmuller for pointing this out.}
$|V_{cb}|$ by $\eta_{\rm QED} = 1 + (\alpha_{\rm em}/\pi) \ln(m_W/\mu) \approx
1.007$. The remarkable agreement between the fit results  shows that the main
difference in the fits is not which short distance $b$ quark mass is used, but
whether $m_b-m_c$ is or is not expanded in terms of HQET matrix elements.

The uncertainties for the \oneSA\ and \kinA\ schemes, which eliminate $m_c$, are
smaller than for the \oneSB\ and \kinB\ schemes, which use
$\overline{m}_c(m_b)$.  This is contrary to the claims made in
\cite{Gambino:2004qm}, but is not unexpected, since the former schemes have only
one parameter at leading order in $1/m_b$, while the latter schemes have two
such parameters. While not expanding in $1/m_c$ gives slightly larger errors
than expanding, the consistency of the central values between the two methods
shows that one can use the $1/m_c$ expansion for inclusive $B$ decays.

One can clearly see that using the kinetic mass for $m_c$ (the
\kinC\ scheme) does not reduce the uncertainties compared to the \oneSA\ and
\kinA\ schemes.  Also, as is now well known, the pole scheme does not work as
well in inclusive calculations as the schemes which use a short distance mass.
Thus, in the remainder of this work we will present results in the  \oneSA\ and
in the \kinA\ schemes. We have carried out the fits in 6 additional schemes,
including the PS and $\overline{\rm MS}$ schemes.  All of the schemes give
reasonable fits, but only the PS scheme with $m_b-m_c$ expanded in HQET gives
rise to similarly small uncertainties as \oneSA\ and \kinA.

The charm quark mass enters into the computation, and we can extract the
value of $m_c$ from our fit. The value of $m_b-m_c$, which is free of the order
$\lqcd$ renormalon ambiguity, is (in the \oneSA\ scheme)
\beq
m_b - m_c =  3.41 \pm 0.01~\GeV\,.
\eeq
We can convert this result to the \MSbar\ mass of the charm quark,
\beqa\label{mcharm}
\mcms(\mcms) &=& 0.90 \pm 0.04~\text{GeV}\,, \nn\\*
\mcms(\mcms) &=& 1.07 \pm 0.04~\text{GeV}\,,
\eeqa
where the two results depend on whether the perturbative conversion factor is
reexpanded or not.\footnote{I.e., the difference between dividing by $1+ a_1
\alpha_s + a_2 \alpha_s^2$ and multiplying by $1-a_1 \alpha_s + (a_1^2-a_2)
\alpha_s^2$. Only the larger value in Eq.~(\ref{mcharm}) has been shown in
Table~\ref{tab:central}.} The reason for the large difference between the two
results is that perturbative corrections are large at the scale $\mcms$. Taking
the average of the two $m_c$ values, and adding half the difference between them
as an additional error gives
\beq
\mcms(\mcms)  = 0.99 \pm 0.1~\text{GeV}\,.
\eeq
The difference between the $m_c$ values  is a nice illustration that one should
avoid using perturbation theory at a low scale, if at all possible. The \kinC\
scheme uses perturbation theory at a scale below $m_c$, and suffers from the
same problem. 

Next, we compare how well the theory can reproduce the experimental
measurements, focusing on the cut dependence of individual moments. The results
for the hadronic moments and the leptonic moments are shown in
Fig.~\ref{fig:hadlepcompare} in the \oneSA\ scheme.  (The DELPHI and CDF results
are included in the fits, but are not shown, as they correspond to $\Ecut=0$.).
The red (dark) shaded band is the uncertainty due to the errors on the fit
parameter. The width of the yellow (light) shaded band is the theoretical
uncertainty due to higher order nonperturbative effects not included in the
computation [the $A$ term in Eq.~(\ref{adef})]. Within the uncertainties, the
OPE predictions for all these moments agree well with the data. As we explained
before, the moments \mX{} and \mX3 were not included in the fit. The yellow
bands shown for \mX{} and \mX3 use $A=0.003$ as an estimate of the uncertainty,
a factor of three larger than for the integer moments, because of the worse
theoretical expansion discussed in  Sec.~\ref{subsec:hadron}.

The agreement between the theory and experiment for the  third lepton moment is
better than our estimate of the theoretical uncertainty. This might be an
indication that we overestimate the theoretical uncertainty for this moment.

The $\chi^2$ for the fit shows that the theory provides an excellent description
of the data. In the \oneSA\ scheme, we get $\chi^2=50.9$ for $\nu=86$ degrees of
freedom, so $\chi^2/\nu=0.59$. The standard deviation for $\chi^2/\nu$ is
$\sqrt{2/\nu}=0.17$, so that  $\chi^2/\nu=0.59$ is about two standard deviations
below the mean value of 1.01. This is some evidence that the theoretical errors
have been overestimated. To study the effect of the theoretical uncertainties,
we also perform fits with all theoretical uncertainties set to zero. This fit
gives $\chi^2 = 148.4$ for the \oneSA\ scheme, and $\chi^2=238.8$ for the \kinA\
scheme. The resulting fits still agree well with the experimental data, as can
be seen from Fig.~\ref{fig:nothcompare}.  The fit results with no theory error
are given in the lower half of Table~\ref{tab:central}. The values
$\chi^2/\nu=1.72$ for the \oneSA\ scheme and $\chi^2/\nu=2.72$ for the \kinA\
scheme are significantly greater than one, which is some evidence that there are
higher order theoretical effects which have not been included.

The calculations in the \kinC\ scheme~\cite{Gambino:2004qm} were used by the
BABAR Collaboration~\cite{Aubert:2004aw}, to perform a fit to its own data. 
While we agree with the results of Ref.~\cite{Gambino:2004qm} for the lepton
energy moments, we are unable to reproduce their results for the hadronic
invariant mass moments. One should also note that Ref.~\cite{Gambino:2004qm} (i)
uses $\alpha_s=0.22$ for the lepton moments, and $\alpha_s=0.3$ for the hadron
moments (ii) includes the $\alpha_s^2 \beta_0$ corrections (which are known for
both the lepton and integer hadron moments) only in the lepton moments, but not
in the hadron moments.

\begin{figure}[t]
\includegraphics[bb=60 190 590 710,height=0.45\textwidth]{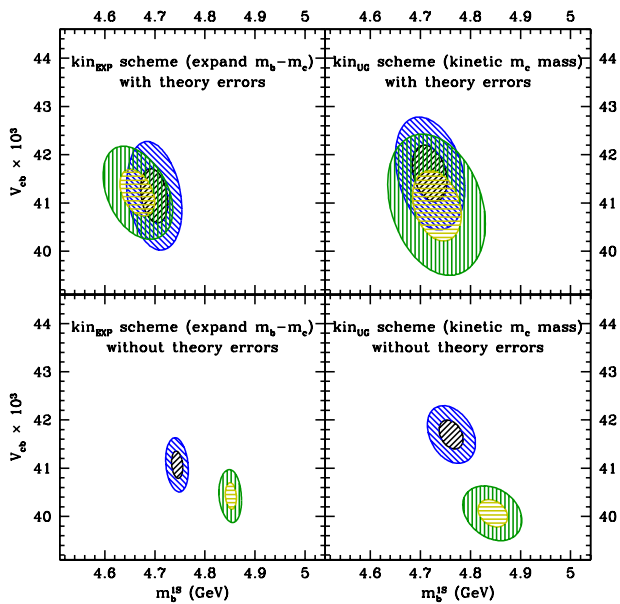}
\caption{Fit results for $|V_{cb}|$ and $m_b$ in the \kinA\ and \kinC\ schemes
using  $\mu_b=1$~GeV (blue and black) and using $\mu_b=1.5\,$~GeV (green and
yellow). $\mu_c$ for the \kinC\ scheme has been kept fixed at 1~GeV. The regions
correspond to $\Delta\chi^2 = 1$ (black and yellow) and 4 (blue and green). The
upper plots includes theory errors in the fit, and the lower plot does not.}
\label{fig:kinCresult}
\end{figure}

\begin{figure}[t]
\includegraphics[bb=60 190 350 710,height=0.45\textwidth]{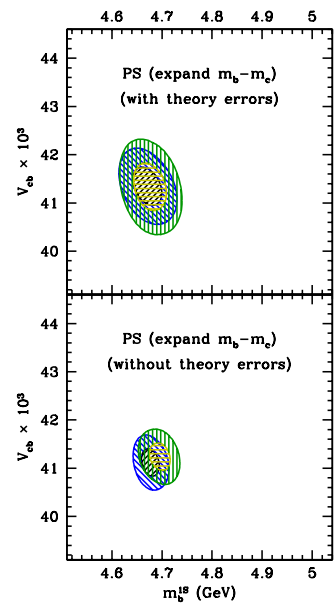}
\caption{Fit results for $|V_{cb}|$ and $m_b$ in the PS scheme using  $\mu =
2\,$~GeV (blue and black) and using $\mu = 1.5\,$~GeV (green and yellow) The
regions correspond to $\Delta\chi^2 = 1$ (black and yellow) and 4 (blue and
green). The upper plots includes theory errors in the fit, and the lower plot
does not.}
\label{fig:ps}
\end{figure}

The \kinA\ and \kinB\ schemes depend on a choice for $\mu_b$.  In the \kinC\
scheme there is an additional dependence on $\mu_c$, and there is no reason why
the theoretical predictions should be expanded using $\mbkin(\mu_b)$ and
$\mckin(\mu_c)$ defined at the same scale ($\mu_b = \mu_c$), since all that is
required is that each $\mu$ should be small, so one has to choose both $\mu_b$
and $\mu_c$. To illustrate the sensitivity to the choice of $\mu_{b,c}$ we show
a fit in Fig.~\ref{fig:kinCresult}  varying $\mu_b$ from 1 to 1.5~GeV keeping 
$\mu_c = 1\,$GeV fixed. Clearly there is significant dependence in the kinetic
schemes with respect to changes in $\mu$, and this should be included as an
additional uncertainty for that scheme. We have included this scale uncertainty
in Table~\ref{tab:central}. The kinetic schemes use perturbation theory at a low
scale $\mu_{b,c}$, and so are sensitive to precisely how these corrections are
included, as was the case for $\mcms(\mcms)$. The PS scheme is much less
sensitive to the value of $\mu$. In Fig.~\ref{fig:ps}, we show the variation
with $\mu$ in the PS scheme. Note that one advantage of the $1S$ scheme is that
it does not depend on any factorization scale parameter $\mu$.

\begin{figure}[tbp]
\includegraphics[bb=60 190 590 710,height=0.45\textwidth]{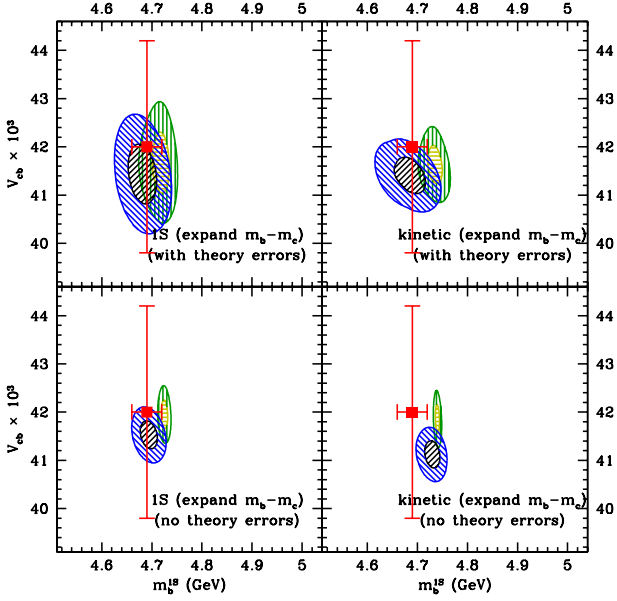}
\caption{Fit results for $|V_{cb}|$ and $m_b$ in the \oneSA\ and \kinA\ schemes.
The upper plots include our estimate of the theoretical errors, the lower ones
set them to zero. The black and blue regions are the fit results ($\Delta
\chi^2=1,4$), and the yellow and green regions are the fit results  ($\Delta
\chi^2=1,4$) omitting $\chi^2_{\text{param}}$ in Eq.~(\ref{delchi}). We have
also shown a red point given by combining Hoang's determination of $m_b^{1S}$
and the PDG 2004 value of $|V_{cb}|$ from exclusive decays.}
\label{fig:oneSresult}
\end{figure}

\begin{table*}[tbp]
$\begin{array}{c|c|c|c|c|c}
\hline\hline
E_{\text{cut}}\ (\text{GeV})  & 0.6 &  0.8 & 1.0 &  1.2 &  1.4  \\
\hline
\displaystyle {{\cal B}(B \to X_c \ell \bar \nu) \over 
  {\cal B}(B \to X_c \ell \bar \nu,E_{\text{cut}})}
= {R_0(0,0) \over R_0(E_{\text{cut}},0)}  & 
1.0523 \pm 0.0001 &   1.1216 \pm 0.0003 & 
1.246 \pm 0.001  & 1.466 \pm 0.001 &  
1.880 \pm 0.003 \\
\hline\hline
\end{array}$
\caption{The ratio of the semileptonic branching ratio ${\cal B}(B \to X_c \ell
\bar \nu)$ to the semileptonic branching ratio with a lepton energy cut
$E_{\text{cut}}$ obtained using the \oneSA\ scheme, including theoretical
uncertainties.\label{tab:slbr}}
\end{table*}

Reference~\cite{Gambino:2004qm} quotes smaller theoretical errors than the
estimates used here, as can be seen from the plots in Ref.~\cite{Aubert:2004aw}. We do not believe that this optimistic estimate of the theoretical uncertainty is justified.

Figure~\ref{fig:oneSresult} shows the results for $|V_{cb}|$ and $m_b$ in the
\oneSA\ and \kinA\ schemes with and without including our estimate of the
theoretical uncertainties. This plot also shows for comparison $m_b=4.69 \pm
0.03\,\GeV$ extracted by Hoang~\cite{hoang} from sum
rules~\cite{Beneke,Voloshin} that fit to the $\bar BB$ system near threshold,
and the PDG 2004 value~\cite{pdg} $|V_{cb}|=\left( 42.0 \pm 1.1 \pm 1.8 \right)
\times 10^{-3}$ from exclusive decays. Hoang's determination of $m_b^{1S}$ is
independent of the current determination, and the agreement is remarkable. The
PDG 2004 value for $|V_{cb}|$ from exclusive decays is also independent of our
determination from inclusive decays.

In summary, we find the following fit results:
\beqa\label{finalresults1}
|V_{cb}| &=& \left( 41.4 \pm 0.6 \pm 0.1_{\tau_B} \right) \times 10^{-3}, \nn\\
m_b^{1S} &=& (4.68 \pm 0.03)\,\text{GeV},
\eeqa
from the \oneSA\ fit including theory errors, where the first error is the
uncertainty from the fit, and the second error (for $|V_{cb}|$) is due to the
uncertainty in the average $B$ lifetime. From the \oneSA\ fit with no theory
errors,  and using the PDG method of scaling the uncertainties so that
$\chi^2/\nu$ is unity, we obtain
\beqa\label{finalresults2}
|V_{cb}| &=& \left( 41.5 \pm 0.4 \pm 0.1_{\tau_B} \right) \times 10^{-3}, \nn\\
m_b^{1S} &=& (4.69 \pm 0.02)\,\text{GeV}.
\eeqa
The increase in $|V_{cb}|$ compared to Ref.~\cite{Bauer:2002sh} is largely due
to an increase in the experimental values for the semileptonic $B$ decay rate
since two years ago.

The \oneSA\ fit (including theoretical uncertainties) also gives
\beq
{\Gamma(B \to X_c \ell \bar \nu) \over |V_{cb}|^2\, \eta_{\text{QED}}^2} = 
(2.49 \pm 0.02) \times 10^{-11} \ \text{GeV} .
\eeq
The ratio of the semileptonic branching ratio with no energy cut to that with an
energy cut is given Table~\ref{tab:slbr}. The semileptonic branching ratio
obtained from the fit (including theoretical uncertainties)  is ${\cal B}(B \to
X_c \ell \bar \nu)  = 0.105 \pm 0.003$. Note that this number depends on the PDG
2004 value (corrected for $B \to X_u$ contamination, see Eq.~(\ref{cont})) of
$0.105\pm0.003$, and the BABAR branching ratio measurements with an energy cut,
which give a higher value of $0.107 \pm 0.002$, when converted to the branching
ratio using Table~\ref{tab:slbr}.

Another useful quantity is the $C$ parameter, needed  for the $B \to X_s \gamma$
rate~\cite{misiak}, which is defined to be
\beq
C = {\Gamma(B \to X_c \ell \bar \nu) \over |V_{cb}|^2}\, 
 {|V_{ub}|^2 \over \Gamma(B \to X_u \ell \bar \nu)} = 0.58 \pm 0.01\,.
\label{Cvalue}
\end{equation}
The value of $C$ depends on the (unknown) matrix element of four-quark
operators, which enter the $B \to X_u \ell \bar \nu$ rate at order $1/m_b^3$
(but not $B \to X_c \ell \bar \nu$).  These absorb a logarithmic divergence in
the $1/m_b^3$ corrections to the $B \to X_c \ell \bar \nu$ rate in the formal
limit $m_c \to 0$. (For $m_c \gg \lqcd$, four-quark operators only enter the $B
\to X_c \ell \bar \nu$ rate at order $\alpha_s$.) The four-quark operator
matrix element gives an uncertainty in addition to that in Eq.~(\ref{Cvalue}).
An estimate of the four-quark operators' contribution is obtained by replacing
the formally divergent term in the $B \to X_u \ell \bar \nu$ rate, $8(
\rho_1/m_b^3) \ln (m_u^2/m_b^2)$~\cite{GK} by $8 (\lqcd /m_b)^3 \sim
0.01$~\cite{BB}.

The above fits give a robust value for $|V_{cb}|$ and $m_b$. However, we
recommend using the error estimate with caution. As we have pointed out, the fit
seems to indicate that the unknown higher order corrections are smaller than our
theoretical estimate of $0.1\%$, so that one can use Eq.~(\ref{finalresults2}).
A theoretical uncertainty less than $0.1\%$ is very small for a hadronic
quantity at the relatively low scale of around 5~GeV.  It is interesting that
the current fit shows that the theoretical uncertainties in inclusive $B$ decay
shape variables are so small. If this is confirmed by further comparisons
between theory and experiment, the uncertainty in $V_{cb}$ can be reduced still
further.

\begin{acknowledgments}

We thank our friends at BABAR, BELLE, CLEO and DELPHI for numerous discussions. We would also like to thank M.~Misiak for pointing out the importance of computing
$C$. This work was supported in part by the US Department of Energy under
Contract DE-FG03-92ER40701 (CWB), DE-AC03-76SF00098 and by a DOE Outstanding
Junior Investigator award (ZL), DE-FG03-97ER40546 (AVM), and  by the Natural
Sciences and Engineering Research Council of Canada (ML and MT).

\end{acknowledgments}

\end{document}